\documentclass[letterpaper,times]{IONconf}

\usepackage{amsmath}
\usepackage{amssymb}
\usepackage{bm}
\usepackage{siunitx}
\usepackage{booktabs}

\usepackage{graphicx}
\usepackage[justification=centering]{caption}
\usepackage{subcaption}
\graphicspath{{figures/}}
\DeclareGraphicsExtensions{.pdf,.jpeg,.png}

\usepackage{array}

\usepackage{url}

\usepackage{natbib} % ion need apalike citations

\usepackage[hidelinks]{hyperref}

\title{Beyond Legacy L1 C/A: Characterizing Advanced Multi-Frequency Spoofing Against Multi-Frequency GPS Receivers for Autonomous Navigation}  
\author{%
  Minhaj Uddin Ahmad,\ Muhammad Sami Irfan,\ Sagar Dasgupta,\ Mizanur Rahman\\
  \small Department of Civil, Construction and Environmental Engineering, The University of Alabama\\[0.3em]
  Thejesh N.\ Bandi\\
  \small Department of Physics and Astronomy, The University of Alabama\\[0.3em]
  \small\texttt{mahmad12@crimson.ua.edu}
}

\begin{document}

\maketitle

% biography section. The * indicates a section excluded from numbering.
\section*{biography}

\biography{Minhaj Uddin Ahmad}{is a PhD student in Transportation Systems Engineering in the Department of Civil, Construction, and Environmental Engineering at the University of Alabama. His research interests include resilient positioning, navigation, and timing (PNT), alternative PNT technologies, sensor fusion, and software-defined radio systems for resilient navigation in GNSS-degraded and -denied environments.}

\biography{Muhammad Sami Irfan}{is a PhD student in Transportation Systems Engineering in the Department of Civil, Construction, and Environmental Engineering at the University of Alabama. His research interests include autonomous vehicle navigation security, GNSS sensor fusion, and resilient navigation systems.}

\biography{Sagar Dasgupta}{is a Research Engineer in Transportation Systems Engineering in the Department of Civil, Construction, and Environmental Engineering at the University of Alabama. His research interests include GNSS-based positioning and navigation security, cyber-physical system security, transportation digital twins, and intelligent transportation systems.}

\biography{Mizanur Rahman}{is an Assistant Professor in Transportation Systems Engineering in the Department of Civil, Construction and Environmental Engineering at the University of Alabama. His research interests include cyber-physical system security and intelligent transportation systems.}

\biography{Thejesh N. Bandi}{is an Associate Professor in the Department of Physics and Astronomy  at the University of Alabama. He is involved in the space clocks R \& D for the Global Navigation Satellite Systems (GNSS), time scales, precision frequency and time synchronization schemes, and related applications.}

\newcommand{\todoX}[1]{\textcolor{red}{[\textbf{TODO:} #1]}}
\newcommand{\citeX}[1]{\textcolor{blue}{[\textbf{CITE:} #1]}}
\newcommand{\statX}[1]{\textcolor{orange}{[\textbf{STAT:} #1]}}
\newcommand{\figX}[1]{\textcolor{violet}{[\textbf{FIG:} #1]}}
\newcommand{\tblX}[1]{\textcolor{violet}{[\textbf{TBL:} #1]}}

\section*{Abstract}

Autonomous vehicles increasingly rely on high-end multi-frequency GNSS receivers that track modernized civil signals, such as L2C, L5, and L1C, alongside legacy GPS L1~C/A. Although multi-frequency operation improves robustness against natural error sources, its resilience against synchronous multi-band spoofing attacks remains insufficiently characterized. This paper presents a method for quantifying the inter-band alignment accuracy required for a coherent spoofer to evade cross-frequency consistency monitoring that a resilient receiver may employ. As a foundational study, the presented method is implemented and evaluated using the GPS L1 C/A and L2C signal pair within a controlled software-defined environment. This environment encompasses deterministic signal synthesis, programmable inter-band code offsets realized through fractional-delay filtering, and processing by a dual-frequency GNSS-SDR receiver. An inter-frequency pseudorange-difference consistency detector is defined and calibrated using a constant-false-alarm-rate (CFAR) approach, yielding a threshold of approximately 19 m at a false-alarm probability of $10^{-3}$. Signal fidelity and receiver capture are verified through correlation-function analysis using a multi-tap correlator bank. The inter-band code offset is then swept from one code chip to 0.005 chips to characterize detector performance. Results show that a perfectly aligned coherent attack yields a detection probability of zero, while detection probability increases rapidly with inter-band misalignment, reaching approximately 50\% at 0.05 chips (about 15 m) and approaching unity for offsets of 0.2 chips or greater. These results quantify the alignment precision required for successful coherent dual-frequency spoofing and establish a practical framework for future evaluation of L5, L1C, and hardware-based multi-frequency spoofing scenarios.

%%%%%%%%%%%%%%%%%%%%%% SECTION %%%%%%%%%%%%%%%%%%%%%%
\section{Background and Motivation}

Modern GNSS receivers increasingly utilize signals beyond legacy GPS L1 C/A. Dual-frequency civil positioning is not itself new; survey-grade receivers have long exploited the L2 band through semi-codeless and codeless tracking of the encrypted P(Y) signal~\citep{woo2000optimum}. However, the codeless tracking approaches are noisy and yield lower signal-to-noise ratio (SNR). The modernized civil L2C signal has changed this landscape as a directly trackable, openly coded signal now broadcast by a large fraction of the constellation~\citep{gpsgov_newcivil}. It has made dual-frequency L1/L2 operation broadly available to commercial off-the-shelf (COTS) receivers. In parallel, the L5 safety-of-life signal continues to expand toward full operational capability. The L1C signal introduced with the GPS Block III satellites, the latest of which was launched in 2026, is intended to provide improved interoperability with other GNSS constellations and to support long-term modernization of the civil service~\citep{lockheed_sv10_2026}. As these signals become standard features of commercial navigation receivers, understanding their security characteristics under spoofing attack becomes increasingly important. Multi-frequency operation improves positioning accuracy by enabling ionosphere-free solutions, enhancing interference resistance, and increasing measurement redundancy~\citep{teunissen2017springer}. As a result, multi-frequency reception is widely regarded as a key component of resilient navigation architectures for safety-critical transportation applications. 

It had been widely suggested that generating coherent counterfeit signals across multiple GNSS bands, with sufficient phase and delay fidelity to satisfy a dual-frequency receiver, was substantially harder and costlier than single-frequency spoofing, and by extension, that multi-frequency survey-grade receivers were largely invulnerable to all but well-funded adversaries~\citep{curran2018infeasibility}. Curran et al. challenged this directly, demonstrating that a low-cost, single-frequency software-defined radio (SDR) could be deliberately misconfigured to produce coherent dual-frequency signals that a survey-grade receiver tracked as a valid L1/L2 position solution, with no measurable code-carrier divergence. This has been done by transmitting a composite signal at a frequency between L1 and L2, such that its harmonics fall at both the L1 and L2 center frequencies. Subsequent work has reinforced the finding that dual-frequency consistency checks, such as Direct Position Comparison, detect single-frequency spoofing but fail against an attacker who reproduces inter-frequency-consistent signals~\citep{wu2025ionospheric}. The feasibility of coherent multi-frequency spoofing is therefore established. What remains uncharacterized is quantitative evaluation. Given that such an attack is possible, how precisely must a spoofer maintain its signals across bands to remain within the tolerance of a receiver's cross-frequency consistency monitoring?

Despite its now-established feasibility, the experimental spoofing literature remains overwhelmingly centered around legacy GPS L1 C/A. Comprehensive reviews of spoofing and anti-spoofing technology~\citep{schmidt2016survey, wu2020spoofing} show that attack demonstrations, benchmark datasets, and defense evaluations are predominantly single-frequency L1 C/A. While the body of research has greatly advanced the understanding of spoofing mechanisms and defenses, it offers limited insight into how modern multi-frequency receivers respond to coherent attacks spanning multiple bands; as a result, many receiver-level threat assessments continue to rest on single-frequency spoofing assumptions that do not reflect the current landscape~\citep{whitehead2018detecting, septentrio_spoofing, broumandan2020nobodys}. The scarcity of multi-frequency studies is driven primarily by the implementation complexity of coherent multi-band signal generation and by the need to maintain consistent code phase, carrier phase, Doppler, and navigation-message timing across widely separated bands, with constraints that tighten further for the higher-chip-rate L5 and L1C signals, rather than by any fundamental infeasibility. These requirements become increasingly demanding for modernized signals, such as L5 and L1C, whose tenfold-higher chip rate imposes substantially tighter timing constraints than legacy signals. 

Rather than treating spoofing detection as a binary outcome, the objective of this study is to quantify the alignment tolerance within which a coherent attack remains undetected and beyond which detection becomes likely. This approach allows researchers to gain insights into the security margin of any detection method and exposes the failure modes of a detection strategy.  Thus, the primary objective of this study is to develop a controlled, repeatable framework to quantify the inter-band alignment accuracy required for successful coherent multi-frequency GNSS spoofing. To this end, we present a controlled software-defined experimental framework based on the GPS L1 C/A and L2C signal pair. The framework manipulates inter-band code delay with sub-chip precision while exposing all relevant receiver observables through a dual-frequency GNSS-SDR~\citep{gnsssdr} software receiver. On this basis, the paper makes three contributions. 

\begin{itemize}
    \item First, an inter-frequency pseudorange-difference consistency detector is defined and calibrated using a constant false-alarm rate (CFAR) criterion. This approach is consistent with how signal-integrity thresholds are usually chosen~\citep{yan2025high}. The calibration employs an explicit quantile threshold, which renders the alignment tolerance re-derivable for any prescribed false-alarm rate. Because this detector operates exclusively on inter-band pseudorange differences and broadcast navigation data, observables provided by virtually all COTS multi-frequency receivers, the resulting performance characterization is directly applicable to COTS units, without requiring access to correlator-level data, which are used in this work solely to validate the signal processing pipeline.
    \item Second, it implements a controlled signal generation and processing pipeline, which is validated via correlation-function analysis employing a multi-tap correlator bank.
    \item Third, to the best of the author’s knowledge, this work presents the first experimentally obtained detection-probability--versus--misalignment characteristic for coherent captured‑receiver spoofing, thereby providing a quantitative measure of the inter-band alignment security margin. Collectively, these results establish a performance benchmark for coherent dual‑frequency spoofing and furnish a methodological basis for assessing L5, L1C, and hardware-based multi-frequency configurations.
\end{itemize}

%%%%%%%%%%%%%%%%%%%%%% SECTION %%%%%%%%%%%%%%%%%%%%%%
\section{Multi-Frequency Spoofing Characterization Framework}\label{sec:framework}

This section provides a controlled and repeatable framework for quantifying the inter-band alignment accuracy required for successful coherent multi-frequency GNSS spoofing. Rather than evaluating spoofing as a binary event, the relative alignment is systematically varied between counterfeit signals transmitted across different frequency bands, and the resulting detection performance of an inter-frequency consistency monitor is measured. This allows the attacker's required synchronization precision to be expressed as a measurable security margin.

\begin{figure}[!htbp]
    \centering
    \includegraphics[width=\linewidth]{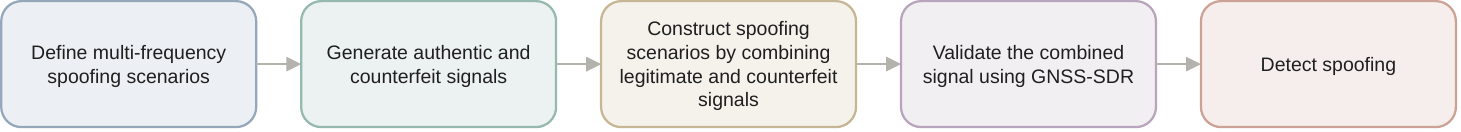}
    \caption{Multi-Frequency Spoofing Characterization Framework}
    \label{fig:framework}
\end{figure}

Figure~\ref{fig:framework} summarizes the overall workflow adopted throughout this study. The framework consists of five sequential stages. First, a coherent multi-frequency spoofing scenario is defined, along with an alignment variable representing the attacker's synchronization error. Second, authentic and counterfeit signals are independently synthesized from broadcast navigation data. Third, the authentic and counterfeit signals are combined to construct controlled spoofing scenarios with configurable inter-band misalignment. Fourth, the combined signals are processed using a dual-frequency software receiver to obtain navigation observables from both frequency bands. Finally, these observables are analyzed using an inter-frequency pseudorange-difference consistency detector to quantify spoofing detectability as a function of inter-band alignment. Each stage of this framework is described in the following subsections.

%%%%%%%%%%%%%%%%%%%%%% Subsection %%%%%%%%%%%%%%%%%%%%%%
\subsection{Multi-frequency spoofing scenarios and assumptions}
The spoofing attack considered in this study is from a coherent dual-frequency attacker capable of generating counterfeit signals simultaneously on the GPS L1 C/A and L2C bands. Unlike conventional single-frequency spoofers, the attacker synthesizes both signals from a common simulated receiver trajectory so that all observables derived from satellite geometry remain mutually consistent. Under this assumption, the spoofed L1 and L2 signals possess correct Doppler relationships, carrier-phase evolution, and navigation-message timing as they originate from the same simulated receiver position and velocity.

The remaining degree of freedom is the relative code alignment between the two frequency bands. This quantity is represented by $\varepsilon=\Delta\tau-\Delta\tau_\mathrm{correct}$ ; where $\tau_\mathrm{correct}$ denotes the correct differential propagation delay between L1 and L2, and $\Delta\tau$ represents the delay reproduced by the spoofer. Consequently, $\varepsilon$ measures the attacker's inter-band synchronization error.

A perfectly synchronized attacker corresponds to $\varepsilon=0$, whereas increasing values of $|\varepsilon|$ represent progressively larger timing errors between the counterfeit L1 and L2 signals. Since all other signal characteristics remain mutually consistent by construction, varying $|\varepsilon|$ isolates inter-band code alignment as the sole experimental variable. This enables the receiver's sensitivity to synchronization errors to be characterized independently of other spoofing artifacts, such as inconsistent Doppler and navigation data.

%%%%%%%%%%%%%%%%%%%%%% Subsection %%%%%%%%%%%%%%%%%%%%%%
\subsection{Signal generation}
The second stage of the framework generates authentic and spoofed GNSS signal datasets using software-defined signal synthesis. Signal generation begins with broadcast ephemeris, which provides satellite positions, velocities, and clock corrections as functions of time. Together with a specified receiver position, these parameters determine the geometric range between each satellite and the receiver. Based on the calculated ranges between the satellites and the receiver, code phase, propagation delay, carrier phase, and Doppler frequency are determined for every visible satellite. These parameters allow the generation of synthetic GPS signals.

The relative satellite-receiver velocity projected along the line of sight determines the Doppler shift experienced on each carrier. Doppler scales directly with carrier frequency; the synthesized L1 and L2 signals preserve the correct inter-frequency Doppler ratio during signal generation. Propagation through the ionosphere is modeled using the broadcast Klobuchar ionospheric model~\citep{klobuchar1987}. Owing to the dispersive nature of the ionosphere, the code delay experienced by each signal depends on carrier frequency according to the inverse-square frequency relationship. Consequently, the synthesized L1 and L2 signals exhibit the expected differential ionospheric delay, which later serves as the reference against which spoofing-induced inconsistencies are measured.

For each receiver trajectory, the contributions from all visible satellites are summed to generate complex baseband samples. Additive noise is included to emulate receiver front-end noise. Two independent signal streams are produced using this procedure. The first represents the authentic signal received by a receiver at its true position. The second represents a counterfeit signal generated for the receiver, starting at the true position and gradually being dragged toward the desired spoofed position. These independently generated data streams provide the basis for constructing controlled spoofing scenarios.

%%%%%%%%%%%%%%%%%%%%%% Subsection %%%%%%%%%%%%%%%%%%%%%%
\subsection{Constructing the spoofing scenario by combining authentic and counterfeit signals} \label{ssec:method-combine}
\begin{figure}[!htbp]
    \centering
    \includegraphics[width=\linewidth]{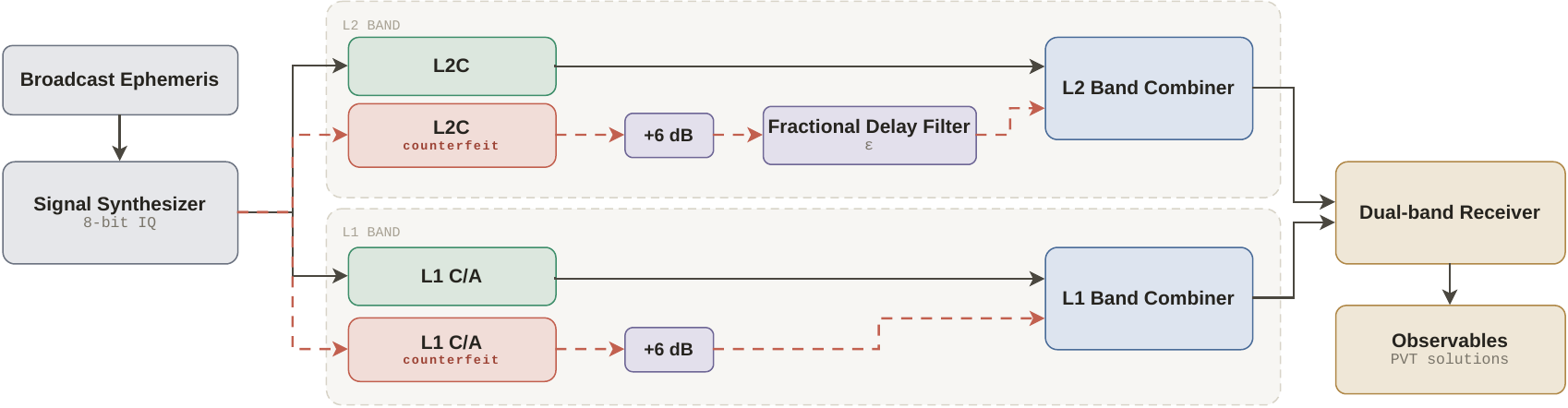}
    \caption{Controlled dual-frequency spoofing signal generation pipeline}
    \label{fig:scenario-gen}
\end{figure}

After generating the authentic and counterfeit signal sets, the two signal streams are combined to construct controlled dual-frequency spoofing scenarios. Figure~\ref{fig:scenario-gen} illustrates the spoofing signal construction pipeline. 

Authentic and counterfeit signals are generated independently for both L1 C/A and L2C. Within each frequency band, the corresponding authentic and counterfeit signals are combined to produce the composite waveform presented to the receiver. The counterfeit signals are transmitted with a constant power advantage of +6 dB relative to the authentic signals. This power margin is selected to ensure reliable receiver capture. The inter-band synchronization error is introduced to the counterfeit L2C signal. A third-order Farrow-structured Lagrange fractional-delay filter applies the desired sub-chip delay prior to combining the counterfeit L2C waveform with its authentic counterpart. Since the authentic signals remain unchanged, the imposed delay perturbs only the relative alignment between the counterfeit L1 and L2 transmissions while preserving all other signal characteristics.

Introducing the synchronization error during waveform synthesis provides precise and repeatable control over the inter-band timing relationship. Unlike coarse transmission-time adjustments, fractional-delay filtering enables continuous variation of chip alignment, allowing the attacker's synchronization accuracy to be characterized over a wide range of operating conditions. The output of this stage consists of synchronized dual-frequency datasets representing authentic reception, perfectly aligned spoofing, and spoofing scenarios with arbitrary inter-band synchronization errors. These datasets are subsequently processed by a dual-frequency software receiver to get navigation observables.

%%%%%%%%%%%%%%%%%%%%%% Subsection %%%%%%%%%%%%%%%%%%%%%%
\subsection{Spoofing signal validation using software receiver} \label{sec:capture}

The composite L1 C/A and L2C datasets are processed using a dual-frequency GNSS-SDR instance configured for L1-aided L2C tracking~\citep{gnsssdr}. For every tracked satellite, the receiver estimates code phase, carrier phase, Doppler frequency, carrier-to-noise ratio, and pseudorange on both frequency bands while simultaneously decoding the navigation message. Since both frequencies share a common receiver clock, clock bias cancels when forming inter-frequency pseudorange differences, allowing raw pseudorange measurements to be directly compared without additional receiver clock estimation.

In addition to navigation observables, GNSS-SDR provides access to internal tracking information, including correlator outputs and tracking loop states. These internal measurements are used solely to verify correct receiver operation and successful spoofing capture during experimental evaluation. The proposed detection algorithm relies solely on pseudorange measurements routinely available from commercial multi-frequency receivers, making the method directly transferable to closed commercial hardware without requiring correlator-level access.

%%%%%%%%%%%%%%%%%%%%%% Subsection %%%%%%%%%%%%%%%%%%%%%%
\subsection{Spoofing detection using inter-frequency pseudorange consistency} \label{sec:charproc}

The final stage of the framework evaluates whether the recovered navigation observables remain consistent across frequency bands. Because ionospheric delay introduces a predictable difference between L1 and L2 pseudoranges, coherent spoofing can be detected by comparing the observed inter-frequency pseudorange difference with its expected value.

For each satellite and measurement epoch, $\Delta\rho_{\mathrm{obs}}(k) = \rho_{L1}(k)-\rho_{L2}(k)$, where $\rho$ denote the measured pseudoranges on the respective frequency bands for each satellite ($k$). The expected differential consists primarily of the dispersive ionospheric delay together with a constant receiver-dependent inter-frequency hardware bias. Rather than estimating these quantities analytically, the expected differential is obtained directly from authentic baseline measurements, thereby capturing both propagation effects and receiver-specific biases within a single reference value. The detector evaluates the absolute deviation, $T_A(k) = \bigl|\Delta\rho_{\mathrm{obs}}(k)-\Delta\rho_{\mathrm{exp}}\bigr|$, which serves as the spoofing detection statistic. Larger values indicate increasing disagreement between the observed and expected inter-frequency relationship. 

To provide a detector whose operating characteristics can be reproduced for different receiver requirements, the decision threshold is determined using a CFAR~\citep{rohling1983radar} criterion. Rather than selecting an arbitrary numerical threshold, the threshold is computed from the empirical distribution of authentic inter-frequency residuals for a specified false-alarm probability. This formulation allows the characterization method to remain independent of any particular receiver implementation while accommodating different operational false-alarm requirements through straightforward recalibration.

The detector therefore produces a binary spoofing decision for every satellite and every measurement epoch. Repeating the complete method over progressively increasing values of the alignment variable $\varepsilon$ yields the relationship between inter-band synchronization accuracy and spoofing detectability, which forms the primary performance metric investigated in this study.

%%%%%%%%%%%%%%%%%%%%%% SECTION %%%%%%%%%%%%%%%%%%%%%%
\section{Evaluation Approach}

This section describes the experimental configuration used to evaluate the proposed characterization framework. A software-defined signal generation pipeline was used to construct authentic and spoofed dual-frequency GNSS datasets under controlled conditions, while a dual-frequency GNSS-SDR receiver processed the resulting signals to recover navigation observables. Before evaluating spoofing detectability, the generated signals were first verified to ensure successful receiver capture and correct tracking behavior on both frequency bands. The presented detector was then evaluated over a series of controlled inter-band alignment offsets to quantify spoofing detection performance.

%%%%%%%%%%%%%%%%%%%%%% Subsection %%%%%%%%%%%%%%%%%%%%%%
\subsection{Experimental platform}
The presented framework was implemented entirely within a software-defined environment to provide complete control over signal generation and inter-band synchronization. Authentic and counterfeit GPS L1 C/A and L2C  signals were synthesized independently using broadcast ephemeris and navigation data corresponding to an authentic receiver located at the University of Alabama, Tuscaloosa ($33.2201^\circ N, -87.5367^\circ W$, approximately 60 m altitude). The simulated observation epoch was set to 24 May 2026, 18:00 UTC, corresponding to a period of relatively high ionospheric activity, ensuring that the naturally occurring differential delay between L1 and L2 remained significant throughout the experiment.

Both frequency bands were generated at a sampling rate of 4.092 MHz using complex 8-bit IQ samples. Since L1 C/A and L2C both have a chipping rate of 1.023 Mcps, this sampling rate provides a fourfold oversampling factor, comfortably exceeding the Nyquist criterion. For each experiment, an authentic signal representing the true receiver position and a counterfeit signal representing a spoofed receiver were generated independently, then combined according to the procedure described in Section~\ref{ssec:method-combine}. Throughout all experiments, the counterfeit signals maintained a constant power advantage of +6 dB relative to the authentic signals to ensure reliable receiver capture.

Signal processing was performed using a single GNSS-SDR instance configured for simultaneous L1 C/A and L2C tracking, with L1-assisted L2 acquisition. Processing both frequency bands within the same receiver instance ensured that all observables were referenced to a common receiver clock and shared tracking architecture, eliminating implementation differences that could otherwise influence inter-frequency comparisons.

%%%%%%%%%%%%%%%%%%%%%% Subsection %%%%%%%%%%%%%%%%%%%%%%
\subsection{Receiver capture validation}  \label{ssec:receiver-val}
Before evaluating spoofing detection performance, it is necessary to verify that the receiver has successfully captured and continuously tracked the counterfeit signals on both frequency bands. The objective of this validation is not to evaluate the detector itself, but to confirm that the generated counterfeit signal can capture the receiver's tracking loop without causing a loss of lock.

To accomplish this, a controlled drag-off scenario was constructed. Initially, the counterfeit signals were aligned with the authentic signals by approximately one code chip, placing both signals within the delay-lock loop pull-in region. The counterfeit signals were then gradually adjusted to steer the receiver solution from the authentic position toward the spoofed position while maintaining continuous tracking. Each drag-off experiment lasted 180 s, allowing sufficient time for initial receiver acquisition, spoofing capture, gradual displacement of the navigation solution, and steady-state operation after capture.

Receiver tracking was evaluated using a multicorrelator correlation-function reconstruction. During processing, correlator outputs recorded by GNSS-SDR were replayed through a bank of delayed code replicas to reconstruct the correlation function over a $\pm 3$-chip delay window for every tracking epoch. Stacking these correlation functions over time produced correlation-function (CAF) waterfalls for both frequency bands.

Figure~\ref{fig:caf} illustrates representative CAF waterfalls for the L1 C/A and L2C signals. In both cases, the tracked spoofed correlation peak remains centered at zero delay while the weaker authentic peak gradually separates as the drag-off progresses. The continuous evolution of the correlation function confirms that receiver lock is successfully transferred to the counterfeit signals without interruption and that simultaneous capture is maintained on both frequency bands throughout the experiment.

%%%%%%%%%%%%%%%%%%%%%% Subsection %%%%%%%%%%%%%%%%%%%%%%
\subsection{Characterization procedure} \label{ssec:charprocedure}
Following successful receiver capture, spoofing detection performance was evaluated under steady-state operating conditions. Unlike the drag-off experiment used for receiver validation, each characterization scenario was generated directly in the spoof-captured state so that measurements reflected only the detector's response to controlled inter-band synchronization errors rather than transient receiver dynamics.

Each experiment consisted of a 120 s dataset, allowing sufficient time for complete navigation-message decoding while providing a long interval of steady-state observations after receiver convergence. To eliminate residual acquisition transients, the first 15 s of each dataset were discarded prior to characterization analysis.

The inter-band synchronization error, represented by the alignment variable $\varepsilon$, was systematically varied over a non-uniform grid spanning perfect synchronization to one complete L1 C/A code chip. Sampling density of the alignment variable was intentionally increased near zero alignment error, where detector sensitivity changes most rapidly, using offsets of 0, $\pm$0.005, $\pm$0.01, $\pm$0.02, and $\pm$0.05 chips. Larger offsets of $\pm$0.1, $\pm$0.2, $\pm$0.5, and $\pm$1 chip were included to characterize detector performance beyond the expected transition region. To improve measurement precision while preserving steady-state behavior, Hatch carrier smoothing~\citep{kim2007adaptive} was applied to both frequency bands using a smoothing window of 100 epochs. The resulting smoothed pseudorange measurements were subsequently used by the inter-frequency consistency detector.

%%%%%%%%%%%%%%%%%%%%%% Subsection %%%%%%%%%%%%%%%%%%%%%%
\subsection{Performance metric}

For each tracked satellite and every measurement epoch, the detector generated an independent spoofing decision by comparing the inter-frequency pseudorange-difference statistic with the calibrated CFAR threshold described in Section~\ref{sec:charproc}. Performance was evaluated using the probability of detection, 

\begin{equation}
    P_D = \frac{N_{detected}}{N_{total}}
    \label{eq:pd}
\end{equation}
where $N_{detected}$ denotes the number of satellite-epoch observations exceeding the decision threshold and $N_{total}$ represents the total number of valid satellite observations collected during the steady-state analysis interval. It is important to note that $P_D$ represents a normalized empirical detection rate. Because the synchronization error remains fixed throughout each experiment, successive observations are temporally correlated through the receiver tracking loops and carrier-smoothing process. The reported detection probability is the fraction of instantaneous satellite observations that are successfully classified during continuous spoofing, as defined in Equation~\ref{eq:pd}. Aggregating measurements across all tracked satellites further provides a better estimate of overall detector sensitivity while naturally accounting for satellite-dependent geometry, measurement noise, and expected inter-frequency pseudorange differences.

%%%%%%%%%%%%%%%%%%%%%% Section %%%%%%%%%%%%%%%%%%%%%%
\section{Experimental Results}
This section presents the experimental evaluation of the presented multi-frequency spoofing characterization framework. We first establish the detector behavior under authentic conditions to calibrate the inter-frequency consistency threshold and verify the expected false-alarm performance. We then evaluate detector sensitivity as the inter-band synchronization error is progressively increased, thereby quantifying the alignment tolerance required for successful coherent spoofing.

%%%%%%%%%%%%%%%%%%%%%% Subsection %%%%%%%%%%%%%%%%%%%%%%
\subsection{Authentic baseline and detector calibration}

The first experiment establishes detector behavior under nominal, non-spoofed operating conditions. Three authentic signal datasets were processed using the dual-frequency receiver described in the previous section to characterize the natural variability of inter-frequency pseudorange differences in the absence of spoofing. These measurements provide the reference statistics required for CFAR threshold calibration.

\begin{figure}[!htbp]
  \centering
  \includegraphics[width=0.9\linewidth]{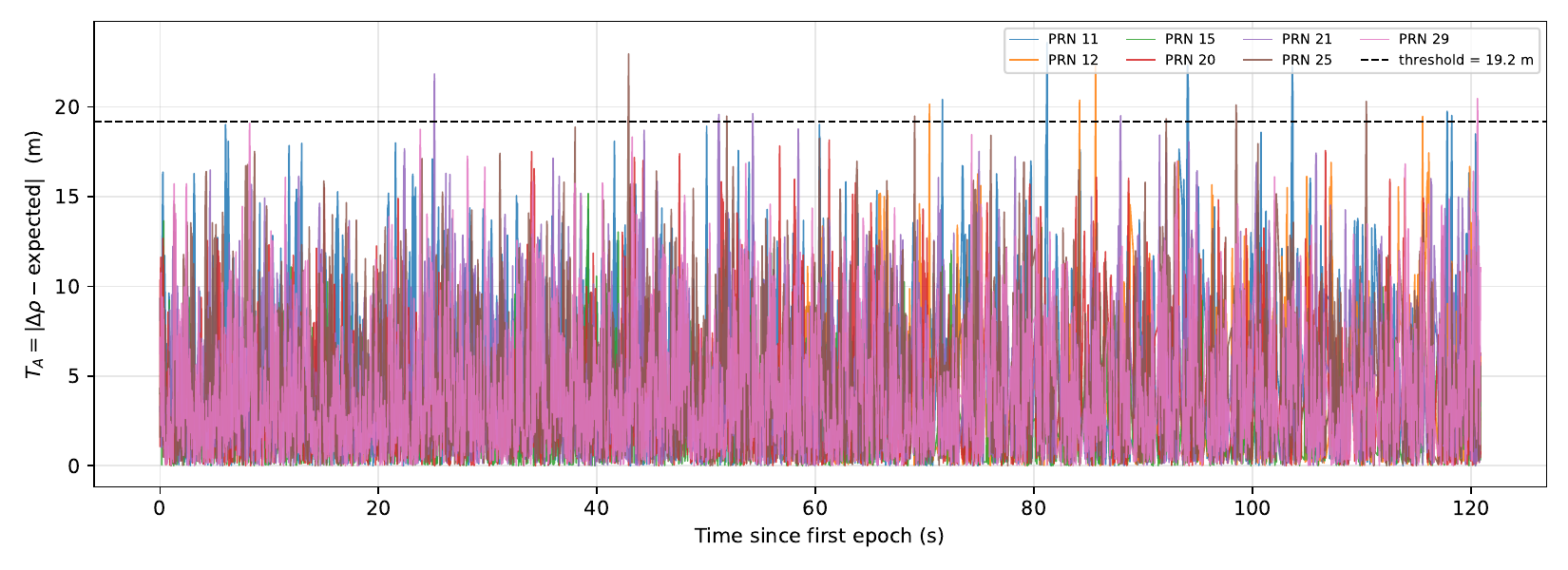}
  \caption{Detector statistic $T_A$ in authentic baseline with a calibrated threshold}
  \label{fig:baseline}
\end{figure}

Figure~\ref{fig:baseline} presents the detector statistic. The inter-band pseudorange difference remains stable for each tracked satellite throughout the observation interval, with individual satellite standard deviations generally ranging between approximately 3 and 7 m. 

Applying the CFAR procedure to the pooled authentic residuals produced a decision threshold of $\gamma_A=19.2$m, corresponding to a target false-alarm probability of $10^{-3}$. Measured detector outputs remain consistently below this threshold throughout the authentic datasets, confirming that the calibrated threshold satisfies the intended false-alarm specification while preserving adequate sensitivity to detect abnormal inter-frequency behavior.

%%%%%%%%%%%%%%%%%%%%%% Subsection %%%%%%%%%%%%%%%%%%%%%%
\subsection{Receiver capture validation}

\begin{figure}[htbp]
  \centering
  \begin{subfigure}{0.49\textwidth}
    \includegraphics[width=\linewidth]{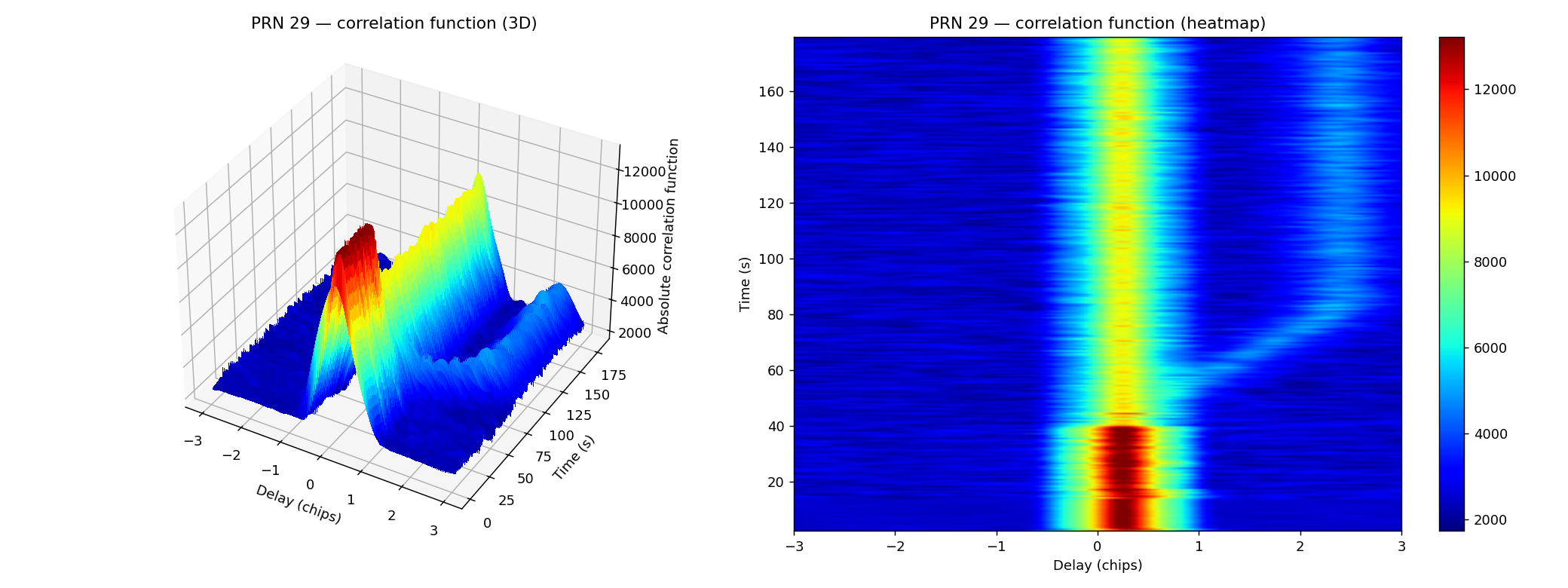}\caption{L1 C/A}
  \end{subfigure}\hfill
  \begin{subfigure}{0.49\textwidth}
    \includegraphics[width=\linewidth]{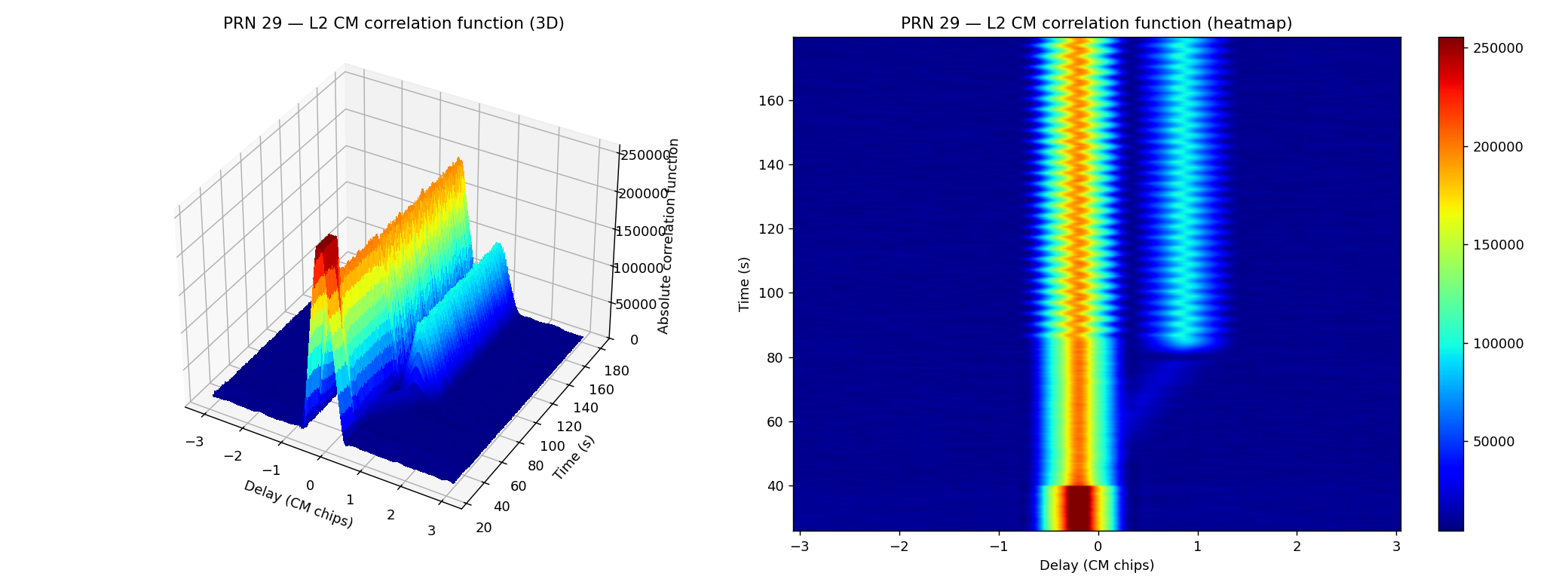}\caption{L2C}
  \end{subfigure}
  \caption{Multicorrelator CAF waterfalls during drag-off capture. The tracked spoof peak is held at zero delay on both bands, while the authentic peak drifts off---validating the synthetic code generation and coherent dual-band capture.}
  \label{fig:caf}
\end{figure}

Before interpreting spoofing detection performance, it is necessary to confirm that the receiver is tracking the counterfeit signals rather than the authentic transmissions throughout the characterization experiments. Figure~\ref{fig:caf} presents representative multicorrelator waterfalls reconstructed from GNSS-SDR tracking outputs during the drag-off experiment described in Section~\ref{ssec:receiver-val}. For both the L1 C/A and L2C signals, the tracked correlation peak remains centered throughout the experiment while the authentic correlation peak progressively separates as the spoofed signal drags the receiver toward the counterfeit position.

These observations verify three important properties of the experimental framework. First, the synthesized GNSS signals preserve the expected correlation characteristics after signal generation and combination. Second, the receiver successfully transfers lock from the authentic signals to the counterfeit signals without interruption. Finally, simultaneous capture is maintained on both frequency bands, ensuring that the subsequent characterization experiments evaluate coherent dual-frequency spoofing under steady-state conditions rather than transient receiver behavior.

Because the proposed detector operates exclusively on navigation observables, establishing successful spoofing capture is essential before interpreting detector performance. The multicorrelator analysis therefore serves as an independent validation of the experimental pipeline rather than as part of the detection algorithm itself.

%%%%%%%%%%%%%%%%%%%%%% Subsection %%%%%%%%%%%%%%%%%%%%%%
\subsection{Alignment-tolerance characterization}

The primary objective of this study is to quantify the relationship between inter-band synchronization accuracy and spoofing detectability. To accomplish this, the synchronization error was systematically varied over the range described in Section~\ref{ssec:charprocedure} while maintaining identical signal generation, receiver configuration, and detector settings for every experiment.

\begin{figure}[htbp]
  \centering
  \includegraphics[width=\linewidth]{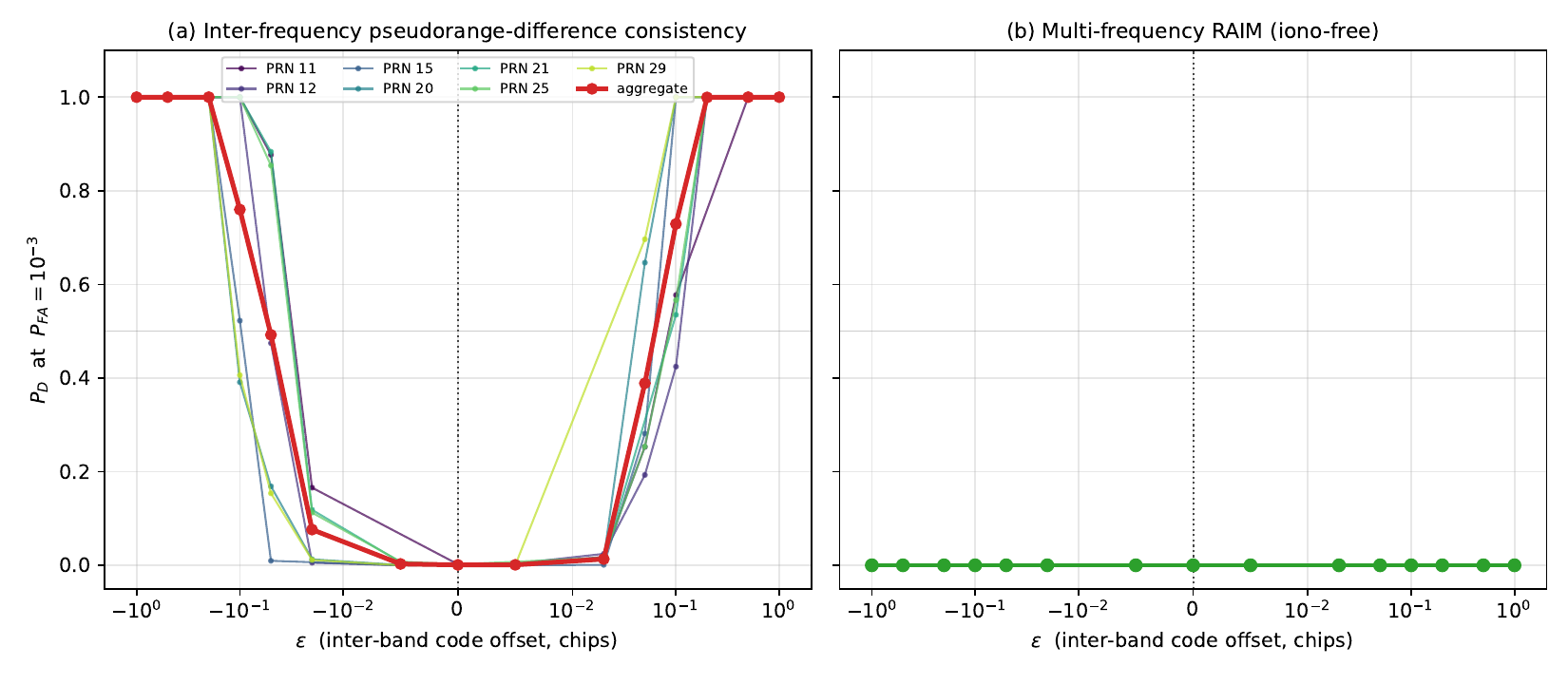}
  \caption{Detection probability vs.\ inter-band offset $\varepsilon$ (symmetric-log axis). The inter-frequency consistency detector (a) shows a sharp well at $\varepsilon=0$ with transition near $\pm0.05$ chips. A multi-frequency-RAIM check (b) is shown for context.}
  \label{fig:sweep}
\end{figure}

Figure~\ref{fig:sweep} summarizes the resulting detection probability as a function of inter-band synchronization error, while the numerical results are reported in Table~\ref{tab:sweep}. The detector exhibits a pronounced minimum at perfect synchronization, followed by a rapid increase in detection probability as the relative timing error increases.

When the counterfeit L1 and L2 signals remain perfectly synchronized ($\varepsilon=0$), the measured detection probability is effectively zero, as shown in Figure~\ref{fig:sweep}(a). Under these conditions, the spoofed signals reproduce the expected inter-frequency pseudorange relationship with sufficient accuracy that the detector fails to distinguish them from authentic satellite measurements. For reference, a multi-frequency Receiver Autonomous Integrity Monitoring (RAIM) detector based on a $\chi^2$ threshold is also implemented and shown in Figure~\ref{fig:sweep}(b) that such coherent attacks are not detected by commonly available RAIM checks. 

For synchronization errors smaller than approximately 0.02 chips, corresponding to roughly 6 m of L1 C/A code delay, detector performance changes only gradually, with detection probabilities remaining below approximately 10\%. This region represents an evasion plateau, where modest synchronization errors remain largely indistinguishable from nominal measurement uncertainty.

\begin{table}[htbp]
  \centering\small
  \caption{Alignment-tolerance characterization: $P_D$ vs.\ $\varepsilon$ (coherent captured-receiver spoofer, $+6$\,dB, $P_{FA}=10^{-3}$).}
  \label{tab:sweep}
  \begin{tabular}{rrl}
    \toprule
    $\varepsilon$ (chips) & $\varepsilon$ (m) & $P_D$ \\
    \midrule
    $0.000$ & $0.0$ & $0.00$ (perfect attack) \\
    $\pm0.005$ & $1.5$ & $\le0.004$ \\
    $\pm0.02$ & $5.9$ & $0.01$--$0.09$ \\
    $\pm0.05$ & $14.7$ & $0.40$--$0.51$ \\
    $\pm0.1$ & $29.3$ & $0.73$--$0.77$ \\
    $\pm0.2$ & $58.6$ & $\approx1.00$ \\
    $\pm0.5,\pm1.0$ & $147,293$ & $1.00$ \\
    \bottomrule
  \end{tabular}
\end{table}

A pronounced transition occurs near an alignment error of approximately 0.05 chips (approximately 15 m). At this point, the detector identifies roughly half of all satellite observations as inconsistent, indicating that the imposed synchronization error has become comparable to the natural variability represented by the calibrated CFAR threshold.

Beyond approximately 0.2 chips, corresponding to approximately 59 m of inter-band code misalignment, detection probability approaches unity. Larger synchronization errors consistently violate the expected physical relationship between L1 and L2 pseudoranges and are therefore detected with near-perfect reliability.

The resulting detection curve demonstrates that spoofing detectability depends primarily on the attacker's ability to maintain extremely precise synchronization between frequency bands rather than simply on transmitting counterfeit signals. The transition region observed around 0.05 chips therefore represents the practical alignment tolerance of the evaluated receiver-detector combination.

\subsection{Perfectly aligned coherent spoofer}

\begin{figure}[!htbp]
  \centering
  \includegraphics[width=0.9\linewidth]{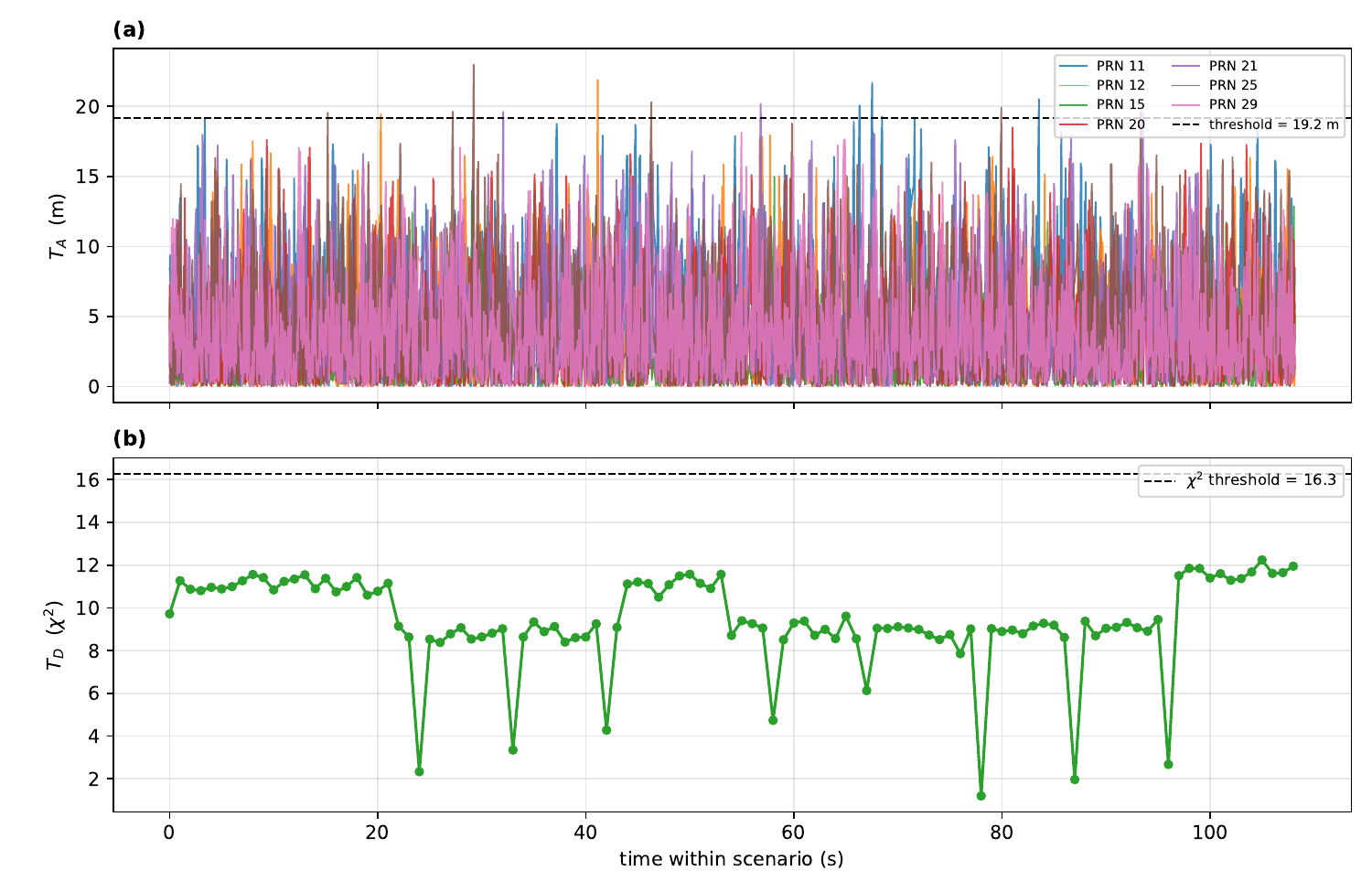}
  \caption{Perfectly aligned coherent attack ($\varepsilon=0$): $T_A$ stays below threshold for all satellites, invisible to the inter-frequency consistency check.}
  \label{fig:tier3}
\end{figure}

Increasing the synchronization error rapidly increases the detection probability. Figure~\ref{fig:tier3} illustrates detector behavior for the $\varepsilon=0$ experiment. Throughout the entire observation interval, the detector statistic remains below the calibrated CFAR threshold for every tracked satellite. No persistent threshold violations are observed despite the counterfeit signals completely capturing the receiver.

This result demonstrates that the proposed detector cannot distinguish a perfectly synchronized, coherent dual-frequency spoofing attack from authentic satellite signals based solely on inter-frequency pseudorange consistency. Rather than indicating a weakness of the detector implementation, this behavior reflects the underlying physics of coherent spoofing. When the attacker faithfully reproduces the expected differential propagation delay between frequency bands, the resulting pseudorange measurements remain physically self-consistent and therefore satisfy the detector's decision criterion.

Consequently, the security margin established in the previous subsection should not be interpreted as a binary detection threshold but rather as the synchronization accuracy that an attacker must achieve to remain undetected. For the receiver configuration evaluated in this study, this corresponds to approximately 0.05 chips, or approximately 15 m of L1 C/A code delay.

%%%%%%%%%%%%%%%%%%%%%% Section %%%%%%%%%%%%%%%%%%%%%%
\section{Discussion and Conclusion}

The experimental results demonstrate that the effectiveness of inter-frequency consistency monitoring depends fundamentally on the attacker's ability to maintain precise synchronization across multiple GNSS frequency bands. Rather than simply transmitting counterfeit signals on L1 and L2, a successful coherent spoofing attack must reproduce the expected differential propagation delay with sufficient accuracy to remain indistinguishable from authentic satellite measurements. For the receiver configuration evaluated in this work, this transition occurs within a relatively narrow synchronization region. Detection probability remains negligible for perfectly aligned signals but increases rapidly once the inter-band synchronization error exceeds approximately 0.05 L1 C/A code chips, corresponding to roughly 15 m of code delay. This threshold represents the practical synchronization accuracy that an attacker must achieve to evade the proposed detector.

These findings provide a different perspective on the security benefits of multi-frequency GNSS. The results presented here indicate that spoofing resistance arises not merely from observing multiple frequencies but from monitoring their consistency. This study also emphasizes the necessity of benchmarking detection methods to expose specific failure modes and define a quantitative security margin within which the detector fails. The proposed characterization framework provides a quantitative approach for evaluating the difficulty of reproducing physically consistent multi-frequency observations for an adversary, independent of any specific detector implementation.

The primary contribution of this study is the establishment of a repeatable method rather than the development of a detector intended to defeat every possible spoofing attack. By introducing controlled inter-band synchronization errors during signal synthesis and systematically measuring detection performance, the framework enables quantification of the synchronization requirements of coherent spoofers using software-defined signal generation and open-source receiver processing. Because the detector operates exclusively on navigation observables that are available from most modern dual-frequency receivers, the presented framework is readily transferable to commercial receiver and may serve as a common evaluation framework for alternative consistency-monitoring algorithms.

Several limitations should be acknowledged when interpreting these results. First, all experiments were conducted within a controlled software-defined environment using synthesized GNSS signals. Although this approach provides precise control over inter-band synchronization and repeatable experimental conditions, it does not capture all impairments present during over-the-air transmission, including antenna effects, front-end distortions, oscillator instabilities, and dynamic radio-frequency propagation. Second, the study considers only GPS L1 C/A and L2C signals under static-receiver conditions with a fixed spoofing-power advantage. The synchronization tolerance observed in this study may differ for other signal combinations, receiver implementations, or dynamic navigation scenarios. Third, the detector relies on empirically calibrated CFAR thresholds derived from authentic baseline observations. While this approach enables robust receiver-specific calibration, threshold values may vary across hardware platforms, environmental conditions, and measurement noise characteristics. 

The presented method naturally lends itself to several directions for future research. The most immediate extension is experimental validation using real-world recorded signals and multi-band RF front ends to determine how hardware impairments influence synchronization tolerance. Extending the framework to additional civil GNSS signals, including GPS L5, Galileo E1/E5, and other modernized multi-frequency constellations, would provide a broader understanding of coherent spoofing across contemporary navigation systems. Dynamic receiver trajectories, realistic vehicular environments, and challenging propagation conditions, such as multipath and signal blockage, should also be incorporated to evaluate detector performance under representative operational scenarios. Finally, although this study focuses on pseudorange consistency, the proposed characterization framework can be generalized to evaluate other multi-frequency observables, including carrier phase, Doppler, and joint consistency metrics. Such extensions would enable systematic comparison of alternative spoofing detection strategies while preserving the central concept introduced in this paper: quantifying the synchronization accuracy required for successful coherent multi-frequency spoofing.

\section*{acknowledgements}

This work is based upon the work supported by the National Center for Transportation Cybersecurity and Resiliency (TraCR) (a U.S. Department of Transportation National University Transportation Center) headquartered at Clemson University, Clemson, South Carolina, USA and National Science Foundation (NSF) (Award \# 2340456). Any opinions, findings, conclusions, and recommendations expressed in this material are those of the author(s) and do not necessarily reflect the views of funding agencies, and the U.S. Government assumes no liability for the contents or use thereof.

We have used generative AI tools for editorial purposes.

% the apacite bibliography style matches the ION bibliography style guidelines.
\bibliographystyle{apalike}
\bibliography{main}

\end{document}